\documentclass[twocolumn,showpacs,preprintnumbers,showkeys,amsmath,amssymb,prl,superscriptaddress]{revtex4}

\usepackage{graphicx}
\usepackage{dcolumn}
\usepackage{bm}

\makeatletter

\newcommand{\Rmnum}[1]{\expandafter\@slowromancap\romannumeral #1@}

\makeatother 	

\begin{document}

\title{Electrodynamic coupling of electric dipole emitters to a fluctuating mode density within a nano-cavity}

\author{Alexey I. Chizhik}
 \email{chizhik@physik3.gwdg.de}
 \affiliation{\Rmnum{3}. Institute of Physics, Georg August University, 37077 G\"ottingen, Germany} 

\author{Ingo Gregor}
 \affiliation{\Rmnum{3}. Institute of Physics, Georg August University, 37077 G\"ottingen, Germany} 

\author{Frank Schleifenbaum}
 \affiliation{Institute of Physical and Theoretical Chemistry, Eberhard Karls University, 72076 T\"ubingen, Germany} 

\author{Claus B. M\"uller}
 \affiliation{\Rmnum{3}. Institute of Physics, Georg August University, 37077 G\"ottingen, Germany} 

\author{Christian R\"oling}
 \affiliation{Accurion GmbH, 37079 G\"ottingen, Germany} 

\author{Alfred J. Meixner}
 \affiliation{Institute of Physical and Theoretical Chemistry, Eberhard Karls University, 72076 T\"ubingen, Germany} 

\author{J\"org Enderlein}
 \email{enderlein@physik3.gwdg.de}
 \affiliation{\Rmnum{3}. Institute of Physics, Georg August University, 37077 G\"ottingen, Germany}

\date{\today}

\begin{abstract}
We investigate the impact of rotational diffusion on the electrodynamic coupling of fluorescent dye molecules (oscillating electric dipoles) to a tunable planar metallic nanocavity. Fast rotational diffusion of the molecules leads to a rapidly fluctuating mode density of the electromagnetic field along the molecules' dipole axis, which significantly changes their coupling to the field as compared to the opposite limit of fixed dipole orientation. We derive a theoretical treatment of the problem and present experimental results for rhodamine 6G molecules in cavities filled with low and high viscosity liquids. The derived theory and presented experimental method is a powerful tool for determining absolute quantum yield values of fluorescence.
\end{abstract}

	

\maketitle

\emph{Introduction}.--- Fluorescing molecules located close to a metal surface (at sub-wavelength distance) or inside a metal nano-cavity, dramatically change their fluorescence emission properties such as fluorescence lifetime, fluorescence quantum yield, emission spectrum, or angular distribution of radiation  \cite{drexhage1974, Kunz1980, Hill2007, Chizhik2009}. This is due to the change local density of modes  of the electromagnetic field caused by the presence of the metal surfaces \cite{Purcell1946}. Although a large amount of studies have dealt with the investigation of this effect, they all have considered fixed dipole orientations of the emitting molecules, so that each molecule exhibits a temporally constant mode density during its de-excitation from the excited to the ground state. However, when  molecules are dissolved in a solvent such as water, their rotational diffusion leads to rapid changes of dipole orientation even on the time-scale of the average excited state lifetime. We will show here that this dramatically influences the coupling of the molecules to the local, strongly orientation-dependent density of modes and the resulting excited state lifetime. This is enormously important for applications of tunable nano-cavities for fluorescence quantum yield measurements. 

\begin{figure}
\centering\includegraphics[keepaspectratio, width=8.5cm]{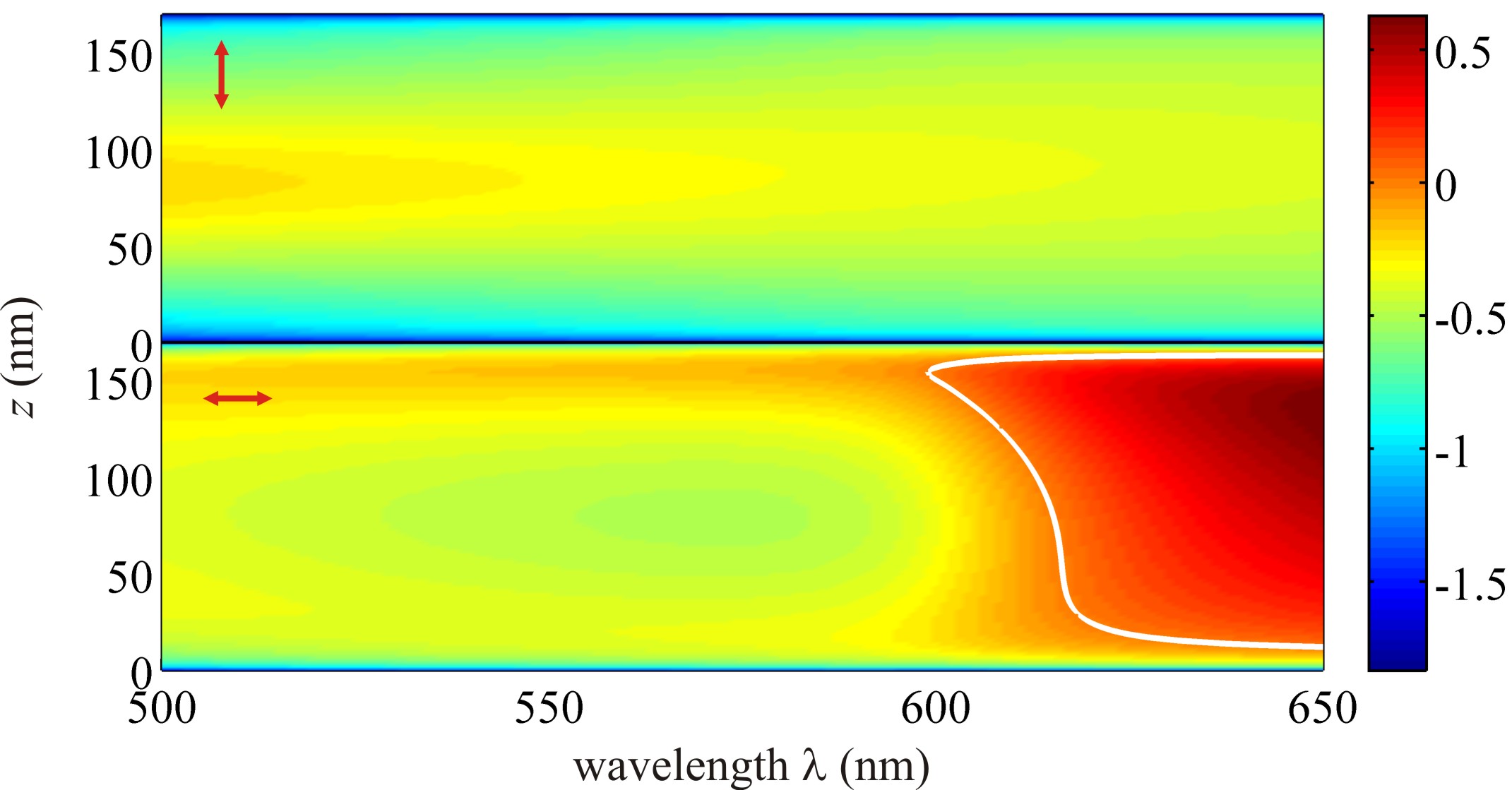}
\caption{\label{fig:rates} Computational results for the decadic logarithm of the relative lifetime, $\lg S_0/S_\mu(\lambda)$, as a function of wavelength $\lambda$ and vertical position $z$ within a nano-cavity comprising of a silver layer of 35 nm (bottom mirror of cavity) and one of 85 nm (top mirror of cavity). The cavity is filled with water. The top panel shows the result for a vertical dipole orientation, the bottom panel for a horizontal orientation. The cavity size, i.e. distance between silver surfaces, is 173~nm corresponding to a white-light maximum of transmission of the cavity of 601~nm. The white line in the bottom panel divides the region where the ratio $S_0/S_\mu(\lambda)$ is larger than one from that where it is smaller than one.}
\end{figure}

\emph{Theory}.---Let us consider an ensemble of molecules within a planar nano-cavity, which had been excited by a short laser pulse into their excited state. Due to the electrodynamic coupling to the cavity, these molecules will exhibit an emission rate $K$ that depends on their vertical position within the cavity, and on the angle $\theta$ between their emission dipole axis and the vertical. In what follows, we assume that the excited state lifetime is so short that one can neglect any translational diffusion of a molecule within the cavity. However, this is in general not the case for its rotational diffusion time which can be on the same order as the excited state lifetime. Then, for a given position within the cavity, the probability density $p(\theta,t)$ to find a molecule still in its excited state at time $t$ with orientation angle $\theta$ obeys the following evolution equation

\begin{equation}
\label{eq:rotdiff}
\begin{split}
\frac{\partial p(\theta, t)}{\partial t} =  \frac{D}{\sin\theta} \frac{\partial}{\partial \theta} \sin\theta \frac{\partial p(\theta,t)}{\partial \theta} - K(\theta) p(\theta,t)
\end{split}
\end{equation}

\noindent where the first term on the right hand side is the rotational diffusion operator \cite{bernepecora} multiplied with rotational diffusion coefficient $D$, and the second term accounts for de-excitation. For the sake of simplicity, we omit any explicit indication of the position dependence of the involved variables. The emission rate $K$ itself is given by a weighted average of the wavelength dependent rates $k(\theta,\lambda)$,

\begin{equation}
K(\theta) = \langle k(\theta,\lambda) \rangle_\lambda = \frac{\int k(\theta,\lambda) F_0(\lambda) d\lambda}{\int F_0(\lambda) d\lambda}
\end{equation}

\noindent where $F_0(\lambda)$ is the free-space emission spectrum of the molecules as a function of wavelength $\lambda$. For a planar cavity, the rates $k$ themselves can be decomposed into

\begin{equation}
k(\theta,\lambda) = k_\perp(\lambda) \cos^2\theta + k_\parallel(\lambda) \sin^2\theta
\end{equation}

\noindent where the $k_{\perp, \parallel}(\lambda)$ are the rates for a vertically and a horizontally oriented emitter, respectively. Within the semi-classical theory of dipole emission \cite{chance1978}, these rates are given by

\begin{equation}
\begin{split}
k_\mu(\lambda)  &= k_{nr} + \frac{S_\mu(\lambda)}{S_0} k_{rad}\\
&= \frac{1}{\tau_0}\left( 1- \Phi + \frac{S_\mu(\lambda)}{S_0}\Phi \right)
\end{split}
\end{equation}

\noindent where the index $\mu$ is either  $\perp$ or $\parallel$,  and where $k_{nr}$ and $k_{rad}$ are the free-space non-radiative and radiative transition rates, respectively, $\tau_0$ is the free-space excited state lifetime, $\Phi$ is the intrinsic quantum yield of fluorescence, $S_\mu(\lambda)$ are the wavelength-dependent emission rates of an oscillating electric dipole with orientation $\mu$ within the cavity, and $S_0$ is the free-space emission rate, which is independent on orientation and wavelength (thus neglecting optical dispersion of the solvent). The emission rates $S_\mu(\lambda)$ are calculated in a semi-classical way by firstly using a plane wave representation of the electromagnetic field of an emitting electric dipole of given orientation (and position) \cite{Girard1996}; secondly calculating the interaction of each plane wave component with the cavity; and finally finding the emission rate as the integral of the Poynting vector of the total field over two surfaces sandwiching the emitter on both sides. An exemplary result for such a calculation is shown in Fig.~\ref{fig:rates}. 

The initial distribution $p(\theta,t=0)$ right after excitation is defined by the polarization and intensity of the focused excitation light. These can be found by again expanding the electromagnetic field of the focused laser beam into a plane wave representation \cite{p6:c26, p6:c27}, and calculating the interaction of each plane wave with the cavity \cite{khoptyar2008}. 
\begin{figure}
\centering\includegraphics[keepaspectratio, width=8.5cm]{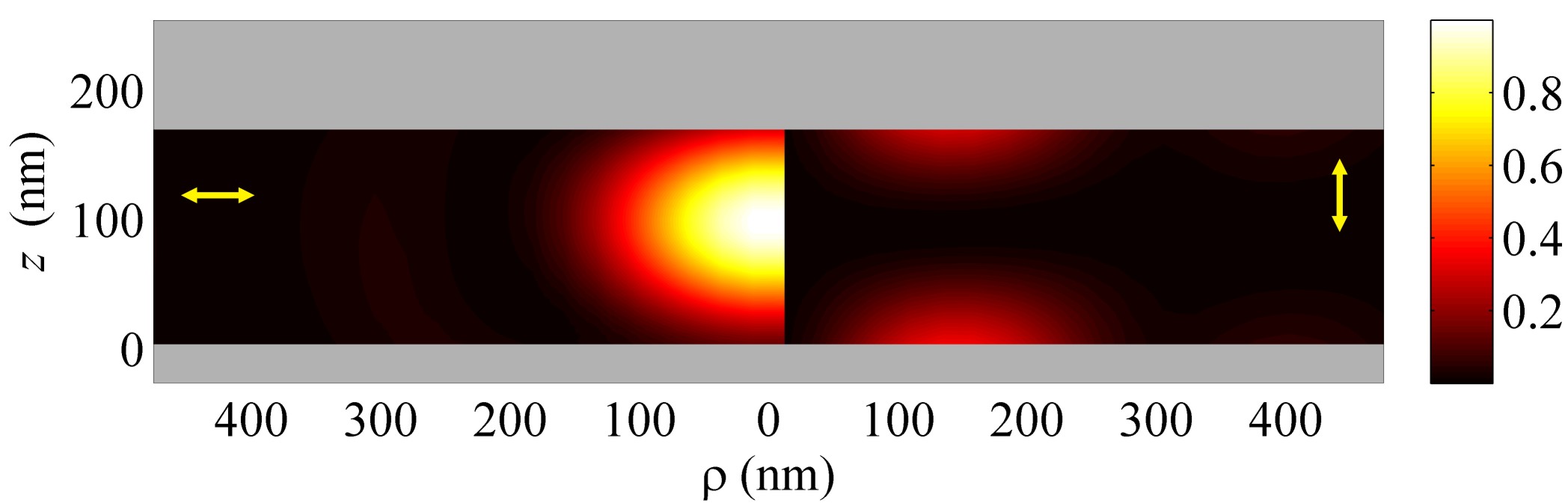}
\caption{\label{fig:exc} Computational results for the normalized excitation intensity distribution within the same nano-cavity as described in Fig.~\ref{fig:rates}. For better visualization of the cavity's geometry, the figure shows also the silver layers as gray-shaded areas. The left side of the figure shows the excitation intensity for horizontally oriented molecules, the right side for vertically oriented molecules.}
\end{figure}
If one denotes the horizontal and vertical components of the excitation intensity at the position of the molecules by $I_\parallel$ and $I_\perp$, respectively, then $p(\theta,t=0)$ is given by 

\begin{equation}
\label{eq:initial}
p(\theta, t = 0) = \frac{3( I_\perp \cos^2\theta + I_\parallel \sin^2\theta)}{2 (I_\perp +2 I_\parallel) }
\end{equation}

\noindent Computational results for $I_\perp$ and $I_\parallel$ are shown in Fig.~\ref{fig:exc}, for the same cavity geometry as in Fig.~\ref{fig:rates}. 

Next, the solution to Eq.~(\ref{eq:rotdiff}) can be found by expanding $p(\theta,t)$ into a series of Legendre polynomials $P_\ell(\cos\theta)$:

\begin{equation}
\label{eq:series}
p(\theta,t) = \sum_{\ell=0}^{\infty}{a_\ell(t)  P_\ell(\cos\theta)}
\end{equation}

\noindent where the $a_\ell(t)$ denote time-dependent expansion coefficients. Inserting that into Eq.~(\ref{eq:rotdiff}) yields an infinite set of ordinary differential equations for the $a_\ell(t)$,

\begin{equation}
\label{eq:al}
\frac{d a_\ell(t)}{d t} = -D \ell (\ell+1) a_\ell(t) - \sum_{\ell^\prime} M_{\ell\ell^\prime} a_{\ell^\prime}(t)
\end{equation}

\noindent where the transition matrix $M_{\ell\ell^\prime}$ is defined by the integrals

\begin{equation}
M_{\ell\ell^\prime} = \frac{2\ell+1}{2} \int_{-1}^1 dx P_\ell(x) P_{\ell^\prime}(x) (x^2 \Delta K + K_\parallel)
\end{equation}

\noindent with the abbreviations $K_{\perp,\parallel} = \langle k_{\perp,\parallel}(\lambda) \rangle_\lambda$, and $\Delta K = K_\perp-K_\parallel$. By carrying out the integration, one finds that the only non-vanishing components of $M_{\ell\ell^\prime}$ are given by

\begin{equation}
M_{\ell\ell^\prime} = \begin{cases}
\frac{(\ell-1)\ell}{(2\ell-3)(2\ell-1)}\Delta K & \mbox{for $\ell^\prime = \ell-2$} \\
\frac{2\ell(\ell+1)-1}{(2\ell-1)(2\ell+3)}\Delta K + K_\parallel & \mbox{for $\ell^\prime = \ell$} \\
\frac{(\ell+1)(\ell+2)}{(2\ell+3)(2\ell+5)}\Delta K & \mbox{for $\ell^\prime = \ell+2$}
\end{cases}
\end{equation}

\noindent From the initial condition, Eq.~(\ref{eq:initial}), one finds that the only non-vanishing initial values $a_\ell$ are

\begin{equation}
\begin{split}
a_0(t=0) = & \frac{1}{2} \\
a_2(t=0) = & \frac{I_\perp - I_\parallel}{I_\perp+2I_\parallel}
\end{split}
\end{equation}

\noindent Although Eq.~(\ref{eq:al}) represents an infinite set of differential equations, it occurs that for our experimental conditions (see below) a truncation of the series expansion of Eq.~(\ref{eq:series}) at a maximum $\ell_{max} = 10$ yields an accurate solution to the problem that does not change when further increasing this truncation value. 

It remains to find an expression for the observable fluorescence emission. This is given by  the integral

\begin{equation}
\label{eq:fluor}
F(t) = \int d\mathbf{r} \int_0^\pi \sin\theta \; p(\theta,t) \langle k(\theta,\lambda) u(\theta,\lambda) \rangle_\lambda 
\end{equation}

\noindent where $u(\theta,\lambda)$ is the orientation and wavelength dependent fluorescence detection efficiency, $\langle \rangle_\lambda$ denotes integration over all wavelengths, and the first integration extends over the whole inner space of the cavity. Due to the rapid fall-off of the excitation intensity when moving a few micrometers away from the center of the focused laser beam, the integration over space can be cut off accordingly. Similarly to the emission rate, the detection efficiency can be represented by 

\begin{equation}
u(\theta,\lambda) = u_\perp(\lambda) \cos^2\theta + u_\parallel(\lambda) \sin^2\theta
\end{equation}

\noindent with $u_\perp$ and $u_\parallel$ being the detection efficiencies for a vertically and horizontally oriented emitter. The most significant cause which makes these detection efficiencies different is the strongly orientation-dependent angular distribution of radiation of the emitters which is collected differently by the detection optics with finite aperture. The detection efficiencies are  calculated again via a plane wave representation of the emitted electromagnetic field, for details see \cite{p6:c40,Enderlein2003}. It should be noted that the detection efficiency goes down to zero when approaching the silver mirrors so that only fluorescence from molecules at least a few nanometers away from the cavity surfaces contributes to the detected signal.

When inserting the expansion (\ref{eq:series}) into Eq.~(\ref{eq:fluor}) and integrating over $\theta$ one finds that only the amplitudes $a_\ell$ with $\ell \in \{0,2,4\}$ contribute to the final result, 

\begin{equation}
F(t) = \int d\mathbf{r} \left[ C_0 a_0(t) + C_2 a_2(t)  + C_4 a_4(t) \right]
\end{equation}

\noindent while the constant factors $C_\ell$ are given by 

\begin{equation}
\begin{split}
C_0 =& \frac{2}{15} \langle 3 k_\perp u_\perp + 2 k_\parallel u_\perp + 2 k_\perp u_\parallel + 8 k_\parallel u_\parallel \rangle_\lambda\\
C_2 =& \frac{4}{105} \langle 6 k_\perp u_\perp + k_\parallel u_\perp + k_\perp u_\parallel - 8 k_\parallel u_\parallel \rangle_\lambda\\
C_4 =& \frac{16}{315} \langle k_\perp u_\perp - k_\parallel u_\perp - k_\perp u_\parallel + k_\parallel u_\parallel \rangle_\lambda
\end{split}
\end{equation}

\noindent Finally, the observable mean fluorescence lifetime $\langle\tau\rangle$ is found as

\begin{equation}
\label{eq:lifetime}
\langle \tau \rangle = \int_0^\infty dt F(t) t \Bigg/ \int_0^\infty dt F(t)
\end{equation}

\emph{Experiment}.---A homemade nano-cavity consists of two silver mirrors with sub-wavelength spacing. The bottom silver mirror (35 nm thick) was prepared by vapor deposition onto commercially available and cleaned microscope glass coverslides (thickness 170 $\mu$m) using an electron beam source (Laybold Univex 350) under high-vacuum conditions ($\sim$$10^{-6}$ mbar). The top silver layer (85 nm thick) was prepared by vapor deposition of silver onto the surface of a plan-convex lens (focal length of 150 mm) under the same conditions. Film thickness was monitored during vapor deposition using an oscillating quartz unit and verified by atomic force microscopy. The complex-valued wavelength-dependent dielectric constants of the silver films were determined by ellipsometry (nanofilm ep3se, Accurion GmbH, G\"ottingen) and subsequently used for all theoretical calculations. The spherical shape of the upper mirror allowed us to reversibly tune the cavity length by retracting from or approaching to the cavity center. It should be noted that within the focal spot of the microscope objective lens the cavity can be considered as a plane-parallel resonator \cite{steiner2005}. For the lifetime measurements, a droplet of a micromolar solution of rhodamine 6G molecules in water or glycerol was embedded between the cavity mirrors. The cavity length was determined by measuring the white light transmission spectrum \cite{steiner2005, Chizhik2011} using a spectrograph (Andor SR 303i) and a CCD camera (Andor iXon DU897 BV), and by fitting the spectra with a standard Fresnel model of transmission through a stack of plan-parallel layers, where the cavity length (distance between silver mirrors) was the only free fit parameter. 

Fluorescence lifetime measurements were performed with a home-built confocal microscope equipped with an objective lens of high numerical aperture (UPLSAPO, 60$\times$, N.A. = 1.2 water immersion, Olympus). A white-light laser system (Fianium SC400-4-80) with a tunable filter (AOTFnC-400.650-TN) served as the excitation source ($\lambda_{\text{exc}}=$~488 nm). The light was reflected by a dichroic mirror (Semrock BrightLine FF484-FDi01) towards the objective, and back-scattered excitation light was blocked with a long pass filter (Semrock EdgeBasic BLP01-488R). Collected fluorescence was focused onto the active area of an avalanche photo diode (PicoQuant $\tau$-SPAD). Data acquisition was accomplished with a multichannel picosecond event timer (PicoQuant HydraHarp 400). Photon arrival times were histogrammed (bin width of 50 ps) for obtaining fluorescence decay curves, and all curves were recorded until reaching $10^4$ counts at the maximum . Finally, the fluorescence decay curves were fitted with a multi-exponential decay model, from which the average excited state lifetime was calculated according to Eq.~(\ref{eq:lifetime}).

\begin{figure} [t]
\centering\includegraphics[keepaspectratio, width=8.5cm]{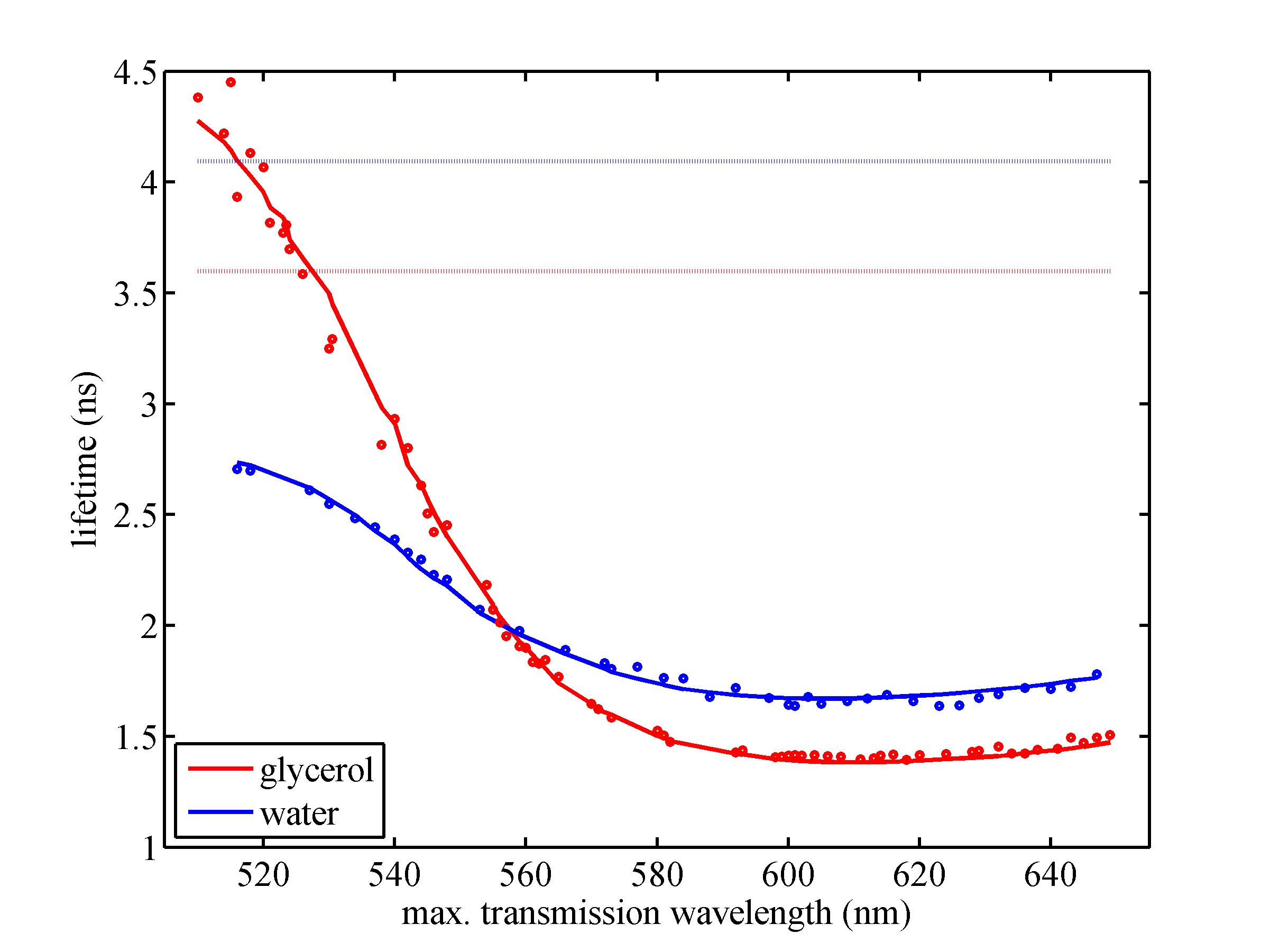}
\caption{\label{fig:lifetime} Fluorescence lifetime of rhodamine 6G in water and in glycerol as a function of the maximum transmission wavelength of the cavity (i.e. cavity size). Dots are experimental data, solid lines theoretical curves, and dotted lines indicate the free-space lifetime values.}
\end{figure}

Fig.~\ref{fig:lifetime} shows the result of the measured average fluorescence lifetime of rhodamine 6G in water (blue dots) and glycerin (red dots) within the nano-cavity as a function of maximum transmission wavelength (which is linearly proportional to cavity length). Both curves show a strong decrease of the lifetime values with increasing cavity length. The solid lines represent fits of the theoretical model to the experimental values, where the only fit parameters have been the free space lifetime $\tau_0$, the fluorescence quantum yield $\Phi$, and the rotational diffusion time $\tau_D = 1/6 D$. For the water solution, the fit values are $\tau_0 = 4.1$~ns, $\Phi = 0.93$, and $\tau_D < 50$~ps, whereas for the glycerol solution they are $\tau_0 = 3.6$~ns, $\Phi = 0.99$, and $\tau_D > 100$~ns (indicating that $D \approx 0$ on the time scale of the fluorescence lifetime). The fluorescence lifetime and fluorescence quantum yield values are in excellent agreement with published values, see \cite{Wurth2011} and citations therein. The large fit value of the rotational diffusion time for rhodamine in glycerol, which is by nearly two orders of magnitude larger than the fluorescence lifetime, indicates that rotational diffusion is practically frozen during de-excitation of the excited molecules, which is similar to the limiting case of fixed dipole orientations. Contrary, the fitted rotational diffusion value in water is significantly shorter than the lifetime, indicating a situation where the emitters perceive an environment with a rapidly fluctuating mode density of the electromagnetic field. Both situations, rapid and slow rotational diffusion, lead not only to quantitatively different results for the dependence of average lifetime on cavity size as seen in Fig.~\ref{fig:lifetime}, but also to qualitatively different behavior: While for slow rotators, the average lifetime can exceed, for specific cavity size values, the free space lifetime (dotted lines in Fig.~\ref{fig:lifetime}), the average lifetime for rapidly rotating molecules will always be smaller than the free-space lifetime. The reason for that can be understood when inspecting Figs.~\ref{fig:rates} and \ref{fig:exc}: The focused laser beam will predominantly excite molecules with horizontal orientation (see Fig.~\ref{fig:exc}), for which the emission rate can be lower than the free-space rate. If the molecules do not rotate, one can thus observe, for specific cavity size values, average lifetime values which are \emph{longer} than the free-space lifetime. However, if molecular rotation is much faster than the average excited state lifetime,  than the emission rate will be dominated by that for vertically oriented molecules (which is much faster than that for horizontally oriented ones, see Fig.~\ref{fig:rates}) and will always result in average lifetime values smaller then the free-space lifetime. Finally, it should be emphasized that the excellent agreement between theoretical model and experimental results offer the fascinating possibility to use lifetime measurements on dye solutions in tunable nano-cavities for simple and direct determination of the fluorescence quantum yield, a quantity which is notoriously difficult to determine by other methods [15]. 

\begin{acknowledgments}
Financial support by the Deutsche Forschungsgemeinschaft is gratefully acknowledged (SFB 937, project A5). 

\end{acknowledgments}

\end{document}